# Fluorescence Phasor Analysis: Basic Principles and Biophysical Applications

Alvaro A. Recoulat Angelini[1,2], Leonel Malacrida[3,4,*], F. Luis González Flecha[,1,2,*]

[1] Laboratorio de Biofísica Molecular, Instituto de Química y Fisicoquímica Biológicas, Universidad de Buenos Aires – CONICET. Buenos Aires, Argentina

[2] Universidad de Buenos Aires – Facultad de Farmacia y Bioquímica. Buenos Aires, Argentina

[3] Unidad de Bioimagenología Avanzada, Institut Pasteur de Montevideo, Hospital de Clínicas. Universidad de la República. Montevideo, Uruguay

[4] Unidad Académica de Fisiopatología, Hospital de Clínicas, Facultad de Medicina, Universidad de la República, Montevideo, Uruguay.

**Abstract**

Fluorescence is one of the most widely used techniques in biological sciences. Its exceptional sensitivity and versatility make it a tool of first choice for quantitative studies in biophysics. The concept of phasors, originally introduced by Charles Steinmetz in the late 19th century for analyzing alternating current circuits, has since found applications across diverse disciplines, including fluorescence spectroscopy. The main idea behind fluorescence phasors was posited by Gregorio Weber in 1981. By analyzing the complementary nature of pulse and phase fluorometry data, he shows that two magnitudes –denoted as G and S– derived from the frequency-domain fluorescence measurements correspond to the real and imaginary part of the Fourier transform of the fluorescence intensity in the time domain. This review provides a historical perspective on how the concept of phasors originates and how it integrates into fluorescence spectroscopy. We discuss their fundamental algebraic properties, which enable intuitive model-free analysis of fluorescence data despite the complexity of the underlying phenomena. Some applications in biophysics illustrate the power of this approach in studying diverse phenomena, including protein folding, protein interactions, phase transitions in lipid mixtures and formation of high-order structures in nucleic acids.

*Keywords:* fluorescence spectroscopy; phasor analysis; Fourier transform; model-free methods

## Historical overview. Gregorio Weber and the birth of biological fluorescence

From the early observations by the Aztecs, who noted a type of wood with medicinal properties that imparted vibrant colors to water, to the first decades of the 20th century—with a pivotal moment in the 1850s through the work of George Gabriel Stokes who described the emission of light by quinine sulfate solutions upon illumination with the "invisible part of sunlight" (Stokes 1852)—fluorescence has attracted the attention of writers, poets and scientists, mainly in the field of physics and, to a lesser and descriptive degree, biology (Valeur and Berberan-Santos 2011; Jameson 2014).

The theoretical foundations of modern fluorescence spectroscopy were firmly established in the first half of the 20th century, after the Quantum theory was formulated. Among the key contributions to this foundational work we can mention: (a) the introduction of energy level diagrams by Jean Perrin and his son Francis Perrin to describe light absorption and emission, later extended to phosphorescence by Alexander Jabłoński (i.e. the well known Perrin-Jablonski diagrams); (b) Fritz Weigert´s observation of partial fluorescence polarization in viscous solvents, leading Francis Perrin to develop a theory of fluorescence polarization; and (c) Enrique Gaviola and Peter Pringsheim's discovery that increasing fluorescein concentration in a viscous solvent reduces fluorescence polarization (explained as a consequence of homotransfer by Francis Perrin) and the subsequent development of a general theory of energy transfer by Theodor Förster (see Jameson 2014; Valeur and Berberan Santos 2012; and references therein).

In the mid-20th century, Gregorio Weber, an Argentine physician born in Buenos Aires, moved to the University of Cambridge, where he began groundbreaking research on the fluorescence of flavins and flavoproteins under the mentorship of Malcolm Dixon. Inspired by the readings of the foundational work of Francis Perrin (Berberan-Santos 2001), Weber's doctoral thesis marked the birth of a new scientific discipline: the quantitative application of fluorescence spectroscopy to study biological systems (Jameson 2001).

Following his time at Cambridge, Weber continued his pioneering work at the University of Sheffield and later at the University of Illinois at Urbana-Champaign. During these years, he made a series of transformative discoveries (Jameson 2001). Among them was his prediction and experimental verification of intrinsic protein fluorescence due to aromatic amino acids (Teale and Weber 1957), the development of the excitation-emission matrix method for resolving contributions from multiple fluorophores (Weber 1960), and the use of resonance energy transfer in protein studies (Weber and Teale 1959).

Weber's contributions extended to the characterization, design and synthesis of fluorescent probes. He synthesized dansyl chloride and characterized its use as a probe of protein hydrodynamics. He also demonstrated that anilino-naphthalene sulfonates (ANS), exhibited intense fluorescence in nonpolar solvents but much weaker fluorescence in water (Weber 1952; Weber and Laurence 1954). Furthermore, he designed and synthesized the fluorescent probe 6-dodecanoyl-2-(dimethylamino) naphthalene (LAURDAN) for studying the phenomenon of dipolar relaxation (Weber and Farris 1979). The spectroscopic properties of this probe when immersed in lipid bilayers, combined with the use of microscopy and fluctuation techniques, make LAURDAN today one of the most valuable fluorescent probes to sense membrane heterogeneity at different temporal and spatial levels (Gunther et al. 2021).

Furthermore, Gregorio Weber made substantial contributions to refining fluorescence lifetime measurements. It should be noted that fluorescence instrumentation was mainly home-built at that time (Jameson 2001). Based on Enrique Gaviola's phase delay technique (Gaviola 1927), Weber developed the "cross-correlation" method in phase fluorometry, which became a foundational technique for modern phase lifetime measurements (Spencer and Weber 1969). This innovative approach is the basis for today's frequency-domain instruments used in fluorescence lifetime analysis.

Gregorio Weber's contributions transformed biological fluorescence from a largely qualitative technique into a precise, quantitative tool that remains indispensable in biological research today. His work has left an indelible mark on biophysics and continues to shape the field of biological fluorescence (Jameson 1998).

**Fluorescence data in the spectral and time domains**

Fluorescence stands as one of the most widely used techniques in biophysics, renowned for its exceptional sensitivity and versatility. A multitude of studies emphasize the significance and broad applications of both steady-state and time-resolved fluorescence in advancing biological research.

Most *in vitro* fluorescence experiments involving biological samples are conducted by recording the steady-state emitted intensity at wavelengths higher than the excitation wavelength. This emission spectrum (Fig 1a), which reflects the probability distribution of the emitted energies, can be characterized by two key parameters: the total fluorescence intensity, which is the cumulative sum of intensities recorded at each wavelength, and the center of spectral mass, defined as the intensity-weighted average of the emitted energies (Jameson 2014). Monitoring changes in these parameters enables researchers to investigate various biological processes, such as protein folding, molecular interactions, conformational changes, and the dynamics of cellular and molecular processes. Additionally, the total fluorescence intensity exhibits the additive property essential for quantitative thermodynamic analysis (Eftink 1994; Moon and Fleming 2011), further enhancing its utility in biophysical research.

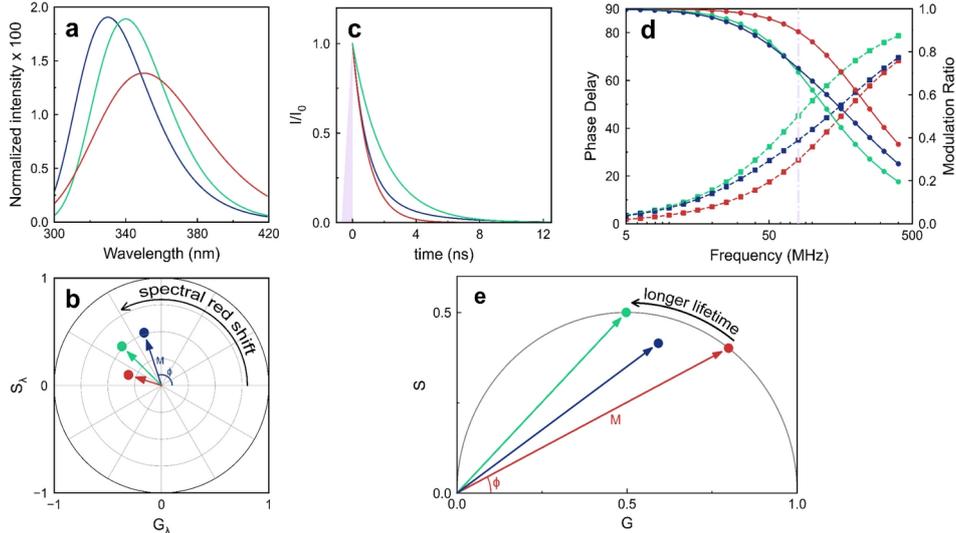

FIGURE 1. Typical spectroscopic fluorescence data and their representation in the phasor space. Three distinct simulated fluorescence emission spectra were normalized using Eq. 10 and plotted in panel a. For each spectrum, a spectral phasor transformation was performed following Eq. 8 and 9, and the results are shown as points in the four-quadrant phasor plot in panel b. Each vector is characterized by a phase angle ($\phi$) and a modulus ($M$). The phase angle is related to the position of the fluorescence spectrum along the wavelength axis, where higher values correspond to spectra with the center of spectral mass at longer wavelengths. The modulus correlates with the width and shape of the spectrum, where higher values indicate narrower spectra. Simulated fluorescence lifetime data of three fluorophores are represented using time-domain (panel c) and frequency-domain (panel d) approaches. Two of them exhibit single-exponential decays, one with a short lifetime (red line) and the other with a longer lifetime (green line), while the third one shows a bi-exponential decay (blue line). Panel e shows the phasor plot for these data, where the time-domain data are processed using Eq. 5-7 and the frequency-domain data are processed with Eq. 3 and 4 using the same frequency (lilac dotted line) as the laser pulse in the time-domain data (lilac region). The phase angle and modulus are indicated in the plot. Monoexponential decays fall on the universal semicircle (gray), with longer lifetimes corresponding to higher phase angles. In contrast, multiexponential decays appear inside the universal semicircle.

Less used, time-resolved fluorescence has enhanced sensitivity, improved specificity, and greater versatility than conventional fluorescence detection methods. In these techniques, data are typically acquired using either time-domain or frequency-domain methodologies. In the time-domain approach, an excitation source delivers a narrow pulse of light (that is shorter than the fluorescence lifetime) and the time course of fluorescence emission is registered (Fig 1c). Modern laser systems are capable of delivering pulses in the picosecond range or even shorter (Liu et al. 2021). Traditional methods for analyzing time-resolved fluorescence involve fitting to the time-dependent fluorescence intensity decay data, of different models, including mono- or multi-exponential decays ,

$$I(t) = \sum_i A_i \cdot e^{-\frac{t}{\tau_i}} \quad (1)$$

where $I(t)$ is the fluorescence intensity at the time $t$, $A_i$ are amplitude coefficients and $\tau_i$ the fluorescence lifetimes. In the case of a monoexponential decay, the lifetime represents the time interval over which fluorescence intensity decreases to ~36.8% ($e^{-1}$) of its initial value. On the other hand, if we consider the emission of photons by an individual molecule, the time between consecutive events follows an exponential (Poisson) distribution with a mean value equal to $\tau$, thus linking the average event rate to the waiting time between events (Lippitz et al. 2005). It is worth noting that in some cases a continuous distribution of lifetimes has to be used to obtain a good fit, as observed when measuring time-resolved Trp fluorescence in proteins (Alcalá et al. 1987).

Conversely, in frequency-domain methods, the sample is excited with a light source whose intensity is sinusoidally modulated at varying frequencies, typically within the megahertz range (Gratton 2016), and the emitted light is registered as a function of time. The emitted fluorescence exhibits a phase shift (θ) and reduced modulation (M) relative to the excitation source. Data obtained from frequency-domain measurements are generally expressed in terms of phase shifts and modulation ratios at each frequency (Fig 1d), and from these data fluorescence lifetimes can be calculated (Ross and Jameson, 2008).

Nowadays, time-resolved fluorescence methods can be classified into two new categories: the phase-modulation shift methods and the time tagging methods (Torrado et al. 2024). The phase-modulation shift methods use modulated light sources for collecting photon arrivals within a given time period. These methods require a modulated laser and specialized electronics. Time tagging methods include time-correlated single photon counting and digital frequency-domain methods, which measure the arrival times of photons corresponding to each laser pulse allowing for accurate measurement of fluorescence decay kinetics and thus the determination of fluorescence lifetimes. The essential components of time tagging instruments include a pulsed laser source, a single photon detector, and electronics capable of time tagging. Among all these techniques, time-correlated single-photon counting has gained prominence given its inherently high sensitivity and timing precision (Farina et al. 2021).

**Charles Steinmetz and the concept of phasors in electrical engineering**

The concept of phasors was introduced by the German-American electrical engineer Charles Proteus Steinmetz as a tool to analyze alternating current (AC) circuits (Araújo and Tonidandel 2013). Steinmetz was a professor and Department chair at Union College in New York, and a pioneering figure in the development of AC systems, which enabled the rapid expansion of the electrical industry in the United States.

Steinmetz's groundbreaking work revolutionized AC circuit theory, which had previously relied on cumbersome, calculus-based methods that were both complicated and time-consuming. By representing sinusoidal currents and voltages with complex numbers, Steinmetz transformed these calculations into a simpler problem of algebra (Steinmetz 1893).

For example, the alternating current intensity was expressed as:

$$I(t) = \sum_i A_i \cdot \cos\omega_i t + j\sum_i A_i \cdot \sin\omega_i t \quad (2)$$

where $I(t)$ is the electrical current intensity at the time $t$, $A_i$ are amplitude coefficients and $\omega_i$ the AC frequencies.

Steinmetz introduced the idea of associating each sinusoidal waveform with a "characteristic circle" and a rotating vector (the phasor), which represented the wave's amplitude and phase angle. This formalization of phasors connected to complex numbers, introduced a powerful way to analyze AC signals. Sinusoidal waveforms can thus be combined by adding or subtracting their horizontal and vertical components in the characteristic circle (i.e. the real and imaginary parts of the complex number), akin to vector addition.

**The phasor transformation of time-resolved and spectral fluorescence data**

In 1981, Gregorio Weber theoretically analyzed the effect on time-resolved fluorescence measurements in the frequency-domain, of the addition of an arbitrary number of sinusoidally modulated components of the same frequency and variable amplitude and phase, and defined two quantities he denoted as $G$ and $S$,

$$G = M \cdot \cos\theta \quad (3)$$

$$S = M \cdot \sin\theta \quad (4)$$

where $\theta$ is the angular delay and $M$ is the relative modulation ratio with respect to the values that characterize the harmonic excitation (Weber 1981).

When analyzing the complementary nature of data from pulse and phase fluorometry, Weber noted that $G$ and $S$ represents the real and imaginary parts of the Fourier transform of the fluorescence intensity:

$$G_\omega = \int_0^\infty I_{norm}(t) \cdot \cos\omega t \, dt \quad (5)$$

$$S_\omega = \int_0^\infty I_{norm}(t) \cdot \sin\omega t \, dt \quad (6)$$

with

$$I_{norm}(t) = \frac{I(t)}{\int_0^\infty I(t)} \quad (7)$$

In these equations $I_{norm}(t)$ is the normalized fluorescence intensity at time $t$ and $\omega$ the angular modulation frequency of the excitation ($\omega = 2\pi/T$ being $T$ the period of the laser pulses).

The similarity between these equations and Eq. 2, which represents the Steinmetz phasor formalism of alternating current intensity, is clear. This analogy was later formalized by two Gregorio Weber's disciples, Enrico Gratton and David Jameson (Jameson et al. 1984; Reinhart et al. 1991; Digman et al. 2008). Fig 1e illustrates the phasors corresponding to the time-resolved fluorescence data shown in Fig 1c. It can be observed that phasors corresponding to monoexponential decays lie on the so-called universal semicircle, while those for multiexponential decays are positioned within the semicircle (Malacrida et al. 2021). These phasor plots also appear in literature under other names, e.g. AB plots (Hanley and Clayton 2005) or polar plots (Redford and Clegg 2005).

In 2012 Fereidouni and co-workers extended the phasor analysis to steady-state fluorescence spectra, as a phenomenological way to analyze the mixed contribution of different fluorescent species (Fereidouni et al. 2012). The spectral phasor components are defined as the real ($G_\lambda$) and imaginary ($S_\lambda$) parts of the Fourier transform of the fluorescence emission spectra:

$$G_\lambda = \sum_\lambda I_{\text{norm}}(\lambda) \cdot \cos\frac{2\pi n(\lambda - \lambda_0)}{L} \quad (8)$$

$$S_\lambda = \sum_\lambda I_{\text{norm}}(\lambda) \cdot \sin\frac{2\pi n(\lambda - \lambda_0)}{L} \quad (9)$$

with

$$I_{\text{norm}}(\lambda) = \frac{I(\lambda)}{\sum_\lambda I(\lambda)} \quad (10)$$

where $I_{\text{norm}}(\lambda)$ are the normalized fluorescence intensity values at each wavelength (i.e. $\Sigma I_{\text{norm}}(\lambda)_i = 1$), $\lambda_0$ is the initial wavelength of the spectrum, $L$ is the length of the registered spectrum ($L = \lambda_f - \lambda_0$) and $n$ is the harmonic value. Fig 1b shows the phasor transformation of the spectra displayed in Fig 1a. Each spectrum is represented by a vector in the phasor space characterized by a modulus (the distance from the origin of the coordinates to the specific point in the phasor space) and a phase angle (the angle between the positive side of the $G_\lambda$-axis and the vector). It was observed that spectral shifts relate with changes in the angular component of the phasor, and spectral widening is related with shortening the phasor modulus. It should be noted that spectral phasors are insensitive to changes in fluorescence intensity (Malacrida et al. 2021). Therefore, if a given process is characterized only by changes in fluorescence intensity, spectral phasor analysis will not provide information about it.

A reinterpretation of the spectral phasors was recently proposed in terms of a complex characteristic function of a probability distribution (the emission spectrum), identifying the argument $2\pi(\lambda - \lambda_0)/L$ as an angular wavelength fraction (Perez Socas and Ambroggio 2022).

**Some properties of phasors**

Phasors can be added through vector addition and multiplied according to standard arithmetic rules for complex numbers (with $j^2 = -1$). Operations derived from these basics include subtraction, division, powers, and roots.

An important mathematical property emerges from the linear combination of two phasors (Weber 1981; Digman et al. 2008; Berberan-Santos 2015; Torrado et al. 2022). A linear combination of vectors involves scaling each vector by a scalar and summing the results (Durrant, 1996). This operation generates a new vector that lies within the span of the original vectors, forming a linear trajectory. In fluorescence phasor analysis, this operation enables the representation of a common natural phenomenon: the two-state processes. In these type of processes fluorescent molecules transition from an initial state (state 1) to a final state (state 2), with each state characterized by a phasor:

$$P_h(1) = G_1 + j \cdot S_1 \quad (11)$$

$$P_h(2) = G_2 + j \cdot S_2 \quad (12)$$

At any stage of the process, a mixture of molecules at the initial and final states is present, and the contribution of each one is given by the relative fraction of photons emitted by the initial state ($x$). The system then evolves from $x = 1$ to $x = 0$. Then, any stage of the process ($i$) will be characterized by the phasor

$$P_h(i) = (G_1 + j \cdot S_1) \cdot x + (G_2 + j \cdot S_2) \cdot (1-x) \quad (13)$$

Applying the rules of the algebra of complex numbers (Cioaba and Linde, 2023), and rearranging the result we obtain

$$P_h(i) = \left[(G_1 - G_2) \cdot x + G_2\right] + j \cdot \left[(S_1 - S_2) \cdot x + S_2\right] \quad (14)$$

As can be seen, both the real component and the imaginary one are linear functions of $x$, and thus the representation of the trajectory in the phasor plane will be characterized by a straight line with slope $(S_2 - S_1)/(G_2 - G_1)$ as shown in Fig 2a.

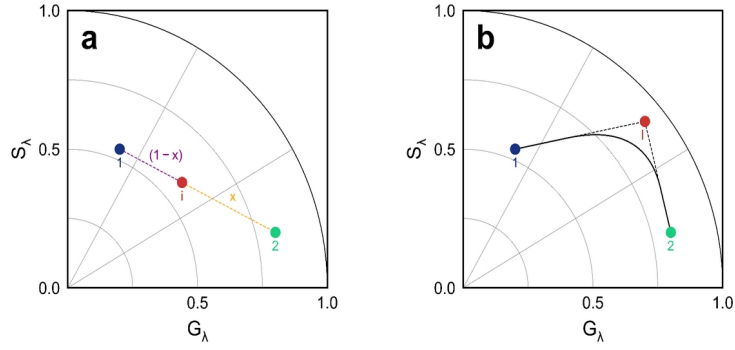

FIGURE 2. Trajectories for two-state and three-state processes on the phasor plot. Panel a shows a two-state transition between state 1 (blue) and state 2 (green), with the transition following a straight line on the phasor plot. As the transition progresses, the phasor point of a given mixture i (red point) will fall along the trajectory. The relative proportions of photons emitted from state 1 (x) and state 2 (1-x) is given by Eq 18. Panel b depicts a three-state transition involving state 1 (blue), an intermediate state I (red), and state 2 (green). The transition will follow straight lines between these states (dotted lines) if each step evolves as a two-states process between each pair of species. However, if the intermediate state is not the dominant species during the transition, its phasor point will be inaccessible, and the trajectory will take a curved shape (solid line).

If the transition under study presents one intermediate species, the process will be represented in a region of the phasor space delimited by the phasors corresponding to each of the three states. In the limiting situation in which the intermediate is the main species at a given stage of the transition, the process will be characterized by two straight lines as shown by dotted lines in Fig. 2b. Otherwise, the trajectory would be non-linear and the vertex corresponding to the intermediate species would not be reached.

Another important property is related to the distance between two points in the phasor space. The distance between two vectors represents their straight-line separation in space. When combined with the concept of linear combinations, the relative distance of an intermediate point along the path connecting two states provides a measure of the proportion each state contributes to generating that point. This distance is defined as the modulus of the vector connecting these states (Danielson, 2003), i.e. the vector resulting from subtracting the corresponding phasors. For the two-states processes discussed above, the distance in the phasor space between the initial and the final state will be

$$dist(P_h(2), P_h(1)) = \|P_h(2) - P_h(1)\| = \sqrt{(G_2 - G_1)^2 + (S_2 - S_1)^2} \quad (15)$$

and the distance between a given stage of the process and the initial state

$$dist(P_h(i), P_h(1)) = \sqrt{([(G_1 - G_2) \cdot x + G_2] - G_1)^2 + ([(S_1 - S_2) \cdot x + S_2] - S_1)^2} \quad (16)$$

because distributive and associative properties hold for the components of the phasors, Eq. 16 can be rearrange as:

$$dist(P_h(i), P_h(1)) = \sqrt{((G_2 - G_1) \cdot (1-x))^2 + ((S_2 - S_1) \cdot (1-x))^2} \quad (17)$$

taking out the common factor $(1-x)^2$ and dividing Eq 17 by Eq 15

$$\frac{dist(P_h(i), P_h(1))}{dist(P_h(2), P_h(1))} = \sqrt{\frac{(1-x)^2 \cdot ((G_2 - G_1)^2 + (S_2 - S_1)^2)}{(G_2 - G_1)^2 + (S_2 - S_1)^2}} = 1 - x \quad (18)$$

This result means that in a two-state process, the relative position of the phasor along the straight line connecting the initial and final states is a measure of the proportion of photons emitted from the final species at this stage of the process. If the process include several intermediate states, the resulting phasor lies within the polygon whose vertices are defined by the phasors of each species, and the distance from the phasor to any of the vertices will be inversely proportional to the fraction of photons emitted by that species (Torrado et al. 2022).

Remarkably, in a real experiment this quantitative information is obtained from the observed phasor trajectory, which results from the photons emitted by each species as mentioned. This doesn't need to assume any model about the underlying mechanisms despite the complexity of the biological and photophysical phenomena involved.

Another important property of phasors emerges when we change the way we observe a given system from analyzing correlated events along a given process (e.g. in the two-state process described above) to analyzing a given system in a single configuration but observing what happens in different parts of the system. This latter mode of analysis is the default mode in fluorescence microscopy and imaging. In this way, we can obtain information from the fluorescence signal (spectral or time-resolved) corresponding to each pixel in the image. With this information, we can calculate the phasor corresponding to each pixel, and we can represent all this information in the phasor diagram. For example, fluorescence lifetimes from various areas of an image can result in phasors segregated into specific areas of the phasor diagram. The phasor and the pixel from which the phasor originated are linked and can be converted to each other. This important property is called the reciprocity principle of the phasor transformation (Ranjit et al. 2018; Malacrida et al. 2021; Torrado et al. 2022).

Adding another layer of analysis, the phasor coordinates calculated from a given pixel are determined by the emission of all fluorescent species present in the part of the sample represented by that pixel, and are related to the phasors of the pure components through the linear combination properties of the phasor space described above, i.e., the phasor position of a pixel with multiple fluorescent species lies within the polygon whose vertices are defined by the phasors of the pure species, and the distance from the image phasor to any of the vertices is inversely proportional to the fraction of photons emitted by that species at that pixel.

**Examples and applications in Biophysics**

*Protein folding*

The three-dimensional native structure of proteins is fundamental to their biological function and is achieved through numerous conformational transitions. Traditionally, protein folding was viewed as a quasi-sequential process involving discrete intermediates along a defined folding pathway, ultimately leading to the native state (Creighton 1988). In contrast, the energy landscape theory offers a more dynamic perspective, describing folding as the gradual organization of a diverse ensemble of partially folded structures that guide the protein toward its final native conformation (Onuchik and Wollynes 2004).

Folding of small globular proteins has been the subject of numerous experimental and computational studies and is now relatively well understood (Finkelstein et al. 2022). Conversely, large multidomain and membrane proteins are much less studied and there is a considerable lack of information about their folding mechanisms (Roman and González Flecha 2014; Rajasekaran and Kaiser 2024).

Fluorescence spectroscopy is a powerful and widely employed technique for investigating protein folding (Eftink 1994; Royer 2006). Among fluorescent probes, tryptophan is the most commonly used due to the remarkable sensitivity of its indole ring's photophysical properties to changes in the local environment. However, when protein folding involves multiple intermediate states, traditional thermodynamic analyses can become intricate. Phasor approaches provide a valuable complementary tool, offering a simplified framework for analyzing complex folding mechanisms (Bader et al. 2014).

James et al. (2011) pioneered the application of in vitro phasor analysis to time-resolved studies on intrinsic protein fluorescence. This approach allowed for the complex decay patterns of protein fluorescence -resulting from multiple emitting tryptophan residues or excited state reactions like energy transfer from tyrosine to tryptophan- to be represented by a single point on a phasor plot. The authors demonstrated that unfolding of lysozyme induced by either urea or guanidine hydrochloride modifies the position of the phasor in a different pattern. Although the aim of that work was not to quantitatively analyse the reversible unfolding of lysozyme, the authors demonstrate that the graphical representation of phasors was a powerful tool to visually identify distinct trajectories, each one indicative of a unique denaturation pathway.

Building on these advances, Bader et al. (2014) employed time-resolved fluorescence, time-resolved fluorescence anisotropy and steady-state fluorescence spectroscopy to analyze the unfolding of a single-tryptophan 179-residue mutant of apoflavodoxin (i.e. flavodoxin without the cofactor FMN) from *Azotobacter vinelandiia*. The analysis of tryptophan fluorescence anisotropy is challenging because decays typically exhibit multiple correlation times. Phasor analysis detected a folding intermediate consistent with prior thermodynamic analysis, without requiring model fitting. Furthermore, assigning a point on the phasor plot to the intermediate state allowed the unmixing of relative proportions of protein states during unfolding, enabling the determination of the free energy values associated with each transition in the unfolding process.

Another excellent tool to explore the unfolding of large multidomain proteins is tryptophan scanning mutagenesis (Vallée-Bélisle and Michnick 2012). Using this approach and the phasor plot method on time-resolved fluorescence, Montecinos-Franjola et al. (2014) analyze the chemical unfolding of the cell division protein FtsZ. This protein is the major cell division protein in bacteria and has emerged as a target for a new generation of antibiotics (Silber et al. 2020). It was observed that, in the native state, the phasor corresponding to each single tryptophan mutant was located at different positions inside the universal semicircle, highlighting distinct environments. Furthermore, different pathways were observed depending on the chemical nature of the denaturing agent and on the position of the single Trp in the structure of FtsZ. The unfolding pathways of the three mutants in urea showed co-localization in the phasor plot at concentrations associated with total secondary structure loss, indicating that it represents the unfolded state. As expected, co-localization of the three mutants in guanidine hydrochloride induced unfolding occurred at lower denaturant concentrations, suggesting that it unfolds FtsZ more effectively.

Expanding on earlier methodologies, variation of the excitation and emission wavelengths and construction of a matrix of the resulting intensities was conceived as a method to elucidate the number of fluorescing compounds in mixtures of fluorophores (Weber 1960). Using this approach Perez Socas and Ambroggio (2022) analyze changes in the phasor representation of this matrix upon urea-induced denaturation of human serum albumin, the most abundant protein in human plasma (Fanali et al. 2012). The authors show that when only the unique tryptophan residue is excited (by excitation at 295 nm), a two-state model is enough to explain the experimental data, but when all the aromatic residues are excited (by excitation at 280 nm) an unfolding intermediate appears. This results in a non-linear trajectory in the phasor representation of the excitation-emission matrix, accounting for a complex scenario for the unfolding of this protein.

While significant progress has been made in the study of folding of small globular proteins, understanding membrane protein folding is one of the major challenges of molecular biophysics. Despite many efforts in recent years, information on protein folding and stability is largely biased towards water-soluble proteins. Recently, chemical denaturation of a Cu(I)-transport ATPase from *Archaeoglobus fulgidus*, induced by the detergent SDS was analyzed by Recoulat Angelini et al. (2024). This thermophilic membrane protein is involved in maintaining $Cu^+$ homeostasis through their active transport across the plasma membrane, coupled to the hydrolysis of ATP (Recoulat Angelini et al. 2021). The SDS-induced changes in Trp fluorescence were analyzed using spectral phasor analysis. The authors observed that three near-linear regions can be identified during the denaturation process, each one corresponding to a two-state event (Fig. 3a). This observation suggested the existence of two intermediate states each characterized by a defined environment of the Trp residues. It is worth noting that traditional thermodynamic analysis only detected the presence of one intermediate analyzing the same data. Dilution of SDS reproduced the same trajectory in the phasor space but with inverse direction, indicating the reversible nature of the SDS induced denaturation process. Spectral phasor analysis was also performed on ANS fluorescence spectra. As was mentioned, the use of this environment-sensitive fluorescent probe in protein studies was pioneered by Gregorio Weber (Weber 1952), and is now established as a molecular marker for molten globule states in protein folding (Ptitsyn 1995). In the case of membrane proteins, background corrected ANS fluorescence mainly reports on the hydrophobic transmembrane region of these proteins (Cattoni et al. 2009; Dodes Traian et al. 2015). Recoulat Angelini et al. reported that the phasor corresponding to the ANS spectra followed a non-linear trajectory in the phasor space, suggesting that no defined intermediate states can be detected at the level of the lipid exposed region of the protein. Taking together these results revealed that SDS-induced denaturation of this protein follows a complex pathway that contrasts with the two-state model derived from previous studies using the same spectroscopic signals, and thermal energy (Cattoni et al. 2008) or guanidine hydrochloride (Roman et al. 2010) as agents for perturbing the native structure. Phasor analysis thus appears as a powerful tool to reflect the structural complexity of this large, multidomain membrane protein.

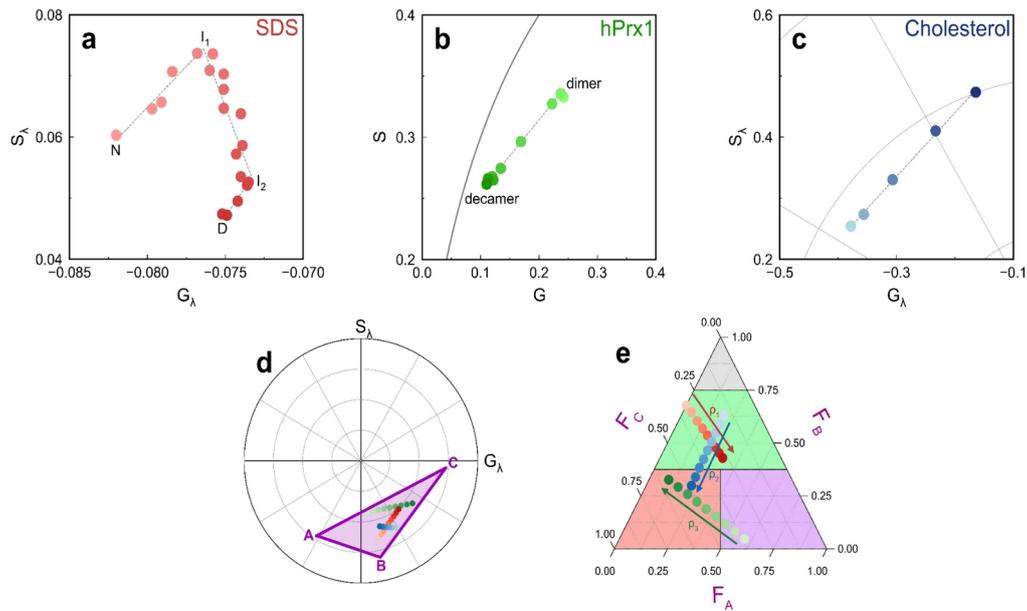

FIGURE 3. Some examples of phasor applications in biophysics. **(a)** Spectral phasor analysis of SDS-induced reversible denaturation of *Archaeoglobus fulgidus* Cu(I)-transport ATPase. The protein was incubated with increasing concentrations of SDS (red points with increasing color intensity), and the Trp emission spectra was recorded and represented on the phasor plot. The denaturation process of the native protein (N) showed three near-linear trajectories, each representing a two-state transition with defined Trp environments, revealing two intermediate states. Adapted from Recoulat Angelini et al. (2024). **(b)** Lifetime phasor analysis of dissociation of human peroxiredoxin 1 (hPrx1). Time-resolved intrinsic fluorescence of hPrx1 at various protein concentrations was recorded and represented using the phasor plot (green points, with increasing color intensity indicating higher protein concentrations). A linear trend is observed with successive dilution of a concentrated sample of hPrx1, linking decamers and dimers. This suggests the absence of intermediate oligomers and enabled the calculation of the decamer-dimer dissociation constant. Adapted from Villar et al. (2022). **(c)** Spectral phasor analysis of phase transitions in 1:1 DOPC:DPPC vesicles by addition of cholesterol. As cholesterol content increased (blue points with increasing color intensity), a linear trend in phasor space reflected a shift in LAURDAN's emission from a "relaxed" state to an "unrelaxed" state, highlighting cholesterol's dehydrating effect on the membrane and its impact on water accessibility around the probe. Adapted from Malacrida et al. (2015). **(d)** LAURDAN's Excitation-Emission Matrices were generated using a computational algorithm based on a model describing LAURDAN's photophysics, involving three probability values, $\rho_1$, $\rho_2$, and $\rho_3$. The absorption and emission of a photon can occur from one of two ground states determined by $\rho_1$. The relaxation process of the excited state is governed by $\rho_2$ or $\rho_3$, depending on the previously excited ground state. This stochastic algorithm generates the resulting excitation-emission spectra, which can then be transformed into phasors using multi-dimensional spectral phasor analysis (Perez Socas and Ambroggio 2022). By fixing two of the probabilities and varying the third ($\rho_1$ in red, $\rho_2$ in blue and $\rho_3$ in green, with increasing color intensity representing higher values), the resulting phasors follow a linear path. By simulating all the probabilities, a triangle is formed by the vertices A ($\rho_1 = 1$, $\rho_3 = 0$), B ($\rho_1 = 0$, $\rho_2 = 0$), and C ($\rho_1 = 0$, $\rho_3 = 1$) which spans the full probability space (purple triangle). Adapted from Perez Soca et al. (2024). **(e)** Ternary diagram for the phasors shown in panel (d). The axes $F_A$, $F_B$, and $F_C$ of the diagram are derived by fractional unmixing each phasor to determine the contributions from the three triangle vertices. A lipid "phase map" was constructed using well-characterized lipid mixtures under various phase-state conditions. The bottom of the triangle represents common non-lamellar self-assembly structures, including "normal" phases (red) and "inverted" phases (purple). The upper section highlights a green area corresponding to lamellar structures. At the top corner lies a "forbidden" region (gray) with no experimental evidence.

### *Protein interactions*

Protein interactions are central to nearly all biochemical processes, serving as the initial step -except in photochemical processes- in driving cellular function (Gutfreund 1995). These interactions play a pivotal role in molecular complex assembly, signal transduction, and enzymatic catalysis. Gregorio Weber works on thermodynamics of ligand-protein interactions under equilibrium conditions are essential for understanding the energetics behind different binding mechanisms (Weber 1992). Furthermore, fluorescence spectroscopy has been fundamental for the experimental study of these interactions, and fluorescence phasors provide a straightforward and model-free approach to analyze binding data, regardless of the complexity of the underlying mechanisms.

A key step in the application of phasor analysis to protein interactions was taken by James et al. (2011). First, they explored the time-resolved Trp fluorescence of dynamin 2. This protein is a GTPase associated with plasma membrane clamping during vesicle fission (Perrais 2022). It is composed of five domains and contains five tryptophan residues. Data analysis using the phasor approach showed that addition of GDP or GTPcS (a slowly hydrolysable GTP analogue), produces a shift in the phasor space that is distinct for each nucleotide. The authors conclude that there is a conformational change around one or more tryptophan residues associated with guanine nucleotide binding. In a second set of experiments, the authors analyzed the fluorescence lifetime of human serum albumin (HSA), in the presence of two drugs: furosemide and D-thyroxine. HSA is one of the most studied proteins due to its ability to bind multiple drugs (Siddiqui et al. 2021). It contains a single Trp residue whose fluorescence has been widely used to study drug binding. The authors showed that in both the studied cases the phasor shifts clockwise relative to free HSA, indicating a shortening of the average lifetime after binding. These changes indicate that drug binding induces changes in the protein conformation near the tryptophan residue and also that the conformational state of the protein differs depending on the nature of the drug. They also observed that the phasor corresponding to dimeric HSA differs from the phasor of monomeric HSA. Finally, the authors also analyzed the effect of mixing two non-interacting proteins, antithrombin and lysozyme, and observed that the phasor corresponding to the mixture was located on the line connecting the phasors of the isolated proteins. On the contrary, when they mixed two proteins whose interaction was well characterized (thrombin/antithrombin), the phasor corresponding to the mixture was located away from the line that connects the phasors corresponding to the two pure proteins.

Subsequent work by Montecinos-Franjola et al. (2014) used time-resolved fluorescence of single Trp mutants as a reporter of conformational changes between the GDP- and GTP-bound states of the cell division protein FtsZ. These mutants were constructed to monitor local conformational transitions related to the nucleotide-bound state at equilibrium. By comparing the phasor representations of the Trp mutants for the GDP- and GTP-bound states, the authors observed that the mutant located at the interface between the N and C domains exhibited the largest conformational change, followed by a slight displacement of the phasor corresponding to mutant located in the nucleotide-binding domain, whereas the mutant located in the catalytic domain reported no change in the phasor space.

Expanding the scope of phasor-based analysis, Pérez Socas and Ambroggio (2020) explored the interaction of a synthetic peptide (21 N-terminal amino acids from the N-terminal matrix domain of the HIV-1 group-specific antigen polyprotein) and liposomes. The presence of a single tryptophan residue in the peptide allowed monitoring the partitioning in membranes since under these conditions a blue shift of the fluorescence emission of tryptophan occurs. This spectral shift was reflected in a shift in the spectral phasor diagram following a straight line, indicating that the insertion of the peptide into the membrane followed a two-state process. Using Eq. (18) the authors calculated the partition constant of the peptide. They also demonstrate that myristoylation of the peptide produced a displacement in the initial position of the phasor, and the trajectory was no longer linear indicating the presence of an intermediate in the insertion of the modified peptide.

Further demonstrating the versatility of phasor analysis, Villar et al. (2022) investigated the dissociation of human peroxiredoxin 1, a decameric protein composed of five head-to-tail homodimers, using fluorescence lifetime and spectral phasors. Peroxiredoxins are a ubiquitous family of multifunctional proteins that are mainly recognized for their antioxidant role, being able to reduce a wide range of hydroperoxides using a specialized cysteine residue (Ferrer Sueta et al. 2011; Villar et al. 2023). The excited-state lifetime and the emission spectra of this protein are different at high and low concentrations, suggesting the presence of two different species. Lifetime phasor analysis showed a straight line connecting the phasors corresponding to dimers and decamers (Fig. 3b), suggesting that the presence of intermediate oligomers is unlikely because either the rapid interconversion between decamers and dimers, or potential intermediates are not stable enough to be detected. A similar observation was made when analyzing the emission spectra using phasor analysis. Once the main trajectory was defined, the authors were able to calculate the fraction of the decamer at each concentration using Eq. (18), and then calculated the decamer-dimer dissociation constant.

### Phase transitions in lipid mixtures

Phase transitions in lipids are fundamental to membrane dynamics and cell function (Shelby and Veatch 2023). These transitions, such as those between gel ($L_\beta$) and fluid ($L_\alpha$) phases in lipid bilayers, are influenced by temperature, hydration, and lipid composition, and they regulate critical membrane properties like fluidity, permeability, and molecular organization. Such changes underpin vital cellular processes, including vesicle fusion, signaling, and membrane protein activity. The ability of lipids to adopt structures like lamellar, hexagonal, and cubic phases, often involves phase transitions that facilitate structural flexibility and functional adaptability. Understanding these transitions is key to deciphering the physical principles governing membrane and cellular behavior.

In their 2011 study, Štefl and coworkers analyze the dipolar relaxation of the solvatochromic probe LAURDAN in DMPC vesicles by examining its emission spectra over a range of temperatures (Štefl et al, 2011). This probe is widely used to study the dynamics and structure of biological membranes, exhibiting distinct emission spectra when incorporated into lipid-water systems that form different lipid phases (Parasassi et al., 1997). Heating DMPC vesicles above the transition temperature shifts LAURDAN's emission maximum from 440 (gel phase) to 480 nm (fluid phase). By measuring time-resolved fluorescence using a 420-nm longpass emission filter, the phasor points move near the universal semicircle, with small deviations outside the circle at higher temperatures. This result suggests that LAURDAN emission, initially from the gel phase is non-relaxed (Parasassi and Gratton 1995) and the increased contribution of LAURDAN molecules in the fluid phase involves excited state reactions such as dipolar solvent relaxation which shift the phasors outside the universal semicircle.

Expanding the use of LAURDAN fluorescence and spectral phasor analysis, Malacrida et al. (2015) investigated the phase coexistence in membranes composed of dioleoyl phosphatidylcholine (DOPC), dipalmitoyl phosphatidylcholine (DPPC) and cholesterol in water, a well-characterized lipid mixture (Veatch and Keller 2003). The study demonstrated that the emission spectrum of LAURDAN in multilamellar vesicles of DOPC in water at room temperature (where membranes are in the fluid phase, $L_\alpha$) shows a maximum at 490 nm. In contrast, multilamellar vesicles of DPPC in water (which form membranes in the gel phase, $L_\beta$, at room temperature) exhibit a spectral maximum at 440 nm. When a 1:1 molar mixture of DOPC and DPPC was used to prepare multilamellar vesicles, LAURDAN's emission spectrum displayed a maximum around 450 nm, indicative of an intermediate state. By converting these spectra into spectral phasors, the authors observed a linear trajectory between the gel and fluid phases, with the 1:1 DOPC:DPPC mixture located at the midpoint. This observation is consistent with the coexistence of fluid-phase and gel-phase membranes within the mixed system. The authors further examined how adding cholesterol to the 1:1 DOPC:DPPC lipid mixture affects the physical behaviour of the vesicles. Cholesterol is known to induce complex structural changes in lipid systems, which are challenging to analyze using traditional methods. However, using the phasor approach, the authors observed a linear trend in the spectral phasor space as the cholesterol content increased from 0% to 50% molar basis (Fig. 3c). This behavior was interpreted in terms of LAURDAN's photophysics, particularly its ability to reorient water molecules around its excited-state dipole at the membrane interface (Parasassi et al. 1997). Under conditions with no water at the membrane interface, LAURDAN exhibits an "unrelaxed" emission. Conversely, when fully surrounded by water molecules (approximately 4–5 water molecules), LAURDAN's emission reflects a "fully relaxed" state, defining two extreme positions in the phasor plot. Intermediate states, involving partial hydration (e.g., 1, 2, or 3 water molecules), correspond to intermediate positions along the linear trajectory in the phasor space. The authors concluded that increasing cholesterol levels in the vesicles progressively reduces the number of water molecules surrounding LAURDAN, thus reflecting a dehydrating effect of cholesterol on the membrane interface.

By combining spectral and time-resolved fluorescence of LAURDAN with phasor analysis, Mangiarotti and Bagatolli (2021) explore water dynamics at the interface of DOPC bilayers in crowded environments. For this study, LAURDAN-labeled multilamellar DOPC vesicles were prepared in the presence of different concentrations of polyethylene glycol (PEG 400). Results showed that, with increasing PEG concentration, the phasors progressively shifted along a trajectory from the fluid-phase ($L_\alpha$) to the gel-phase ($L_\beta$), suggesting a reduced dynamics of the interfacial water. This shift was attributed to a more rigid lipid packing induced by PEG. The study also found that decreased dipolar relaxation of water at the membrane interface correlated with an increased proportion of randomly oriented configurations of polymers. Moreover, the authors demonstrated that protein structural transitions from globular to extended conformations could induce transitions between lamellar and non-lamellar lipid phases. These findings highlight the critical role of water in coupling structural changes in macromolecules with the supramolecular organization of lipids.

Extending the applications of LAURDAN phasor analysis, Pérez Socas et al. (2024) analyzed the excitation-emission matrices in various lipid structures and applied spectral phasor analysis to interpret the data. The authors proposed a theoretical model based on LAURDAN photophysics, enabling the construction of excitation-emission matrices that can subsequently be transformed in a spectral phasor. Their findings revealed that all the phasors are confined to a characteristic region within the phasor space delimited by 3 vertices (Fig. 3d). Each phasor can thus be described as a linear combination of the vertices, and the whole simulated conditions lead to a triangular phase diagram (Fig. 3e). To test their theoretical framework, the authors performed experiments under several well-established lipid conditions. These included studying phospholipid bilayers in different phases, examining cholesterol's effects on bilayer organization, and characterizing both "normal" and "inverted" non-lamellar structures. They also analyzed transitions from non-lamellar to lamellar structures driven by changes in lipid composition and lamellar-to-non-lamellar transformations triggered by phospholipase activity. Their findings validated the model, demonstrating that different phases are segregated in specific regions of the triangular diagram.

*Formation of high-order structures in nucleic acids*

The discovery of the double-helix structure of DNA by Rosalind Franklin, James Watson and Francis Crick marks a landmark moment in our understanding of genetic material. However, it is now increasingly recognized that high-order nucleic acid structures, such as hairpins, triplexes, and G-quadruplexes, also play significant biological roles. G-quadruplexes are high-order DNA or RNA structures found within cells, believed to play essential roles in various cellular processes, including gene regulation, transcription, translation, recombination, and DNA repair (Monsen et al. 2022). These structures are four-stranded helical motifs that can form either intermolecularly or intramolecularly, and can be classified into tetramolecular (four-strand), bimolecular (two-strand), and unimolecular (single-strand) forms, with the latter capable of forming in sequences containing stretches of guanine bases separated by loops (Kolesnikova and Curtis 2019). The core structure consists of stacked tetrads, creating a central cavity that coordinates monovalent cations, such as $K^+$ or $Na^+$, which stabilize the quadruplex structure.

Time-resolved fluorescence techniques provide a powerful tool to explore the wide array of conformations adopted by G-quadruplexes, particularly in human telomeres (Minsky 2004). A commonly used fluorescent adenine analog, 2-aminopurine (2AP), is valuable for probing local DNA environments (Neely et al. 2005). However, 2AP exhibits a multi-exponential fluorescence lifetime when incorporated into DNA oligomers, making it challenging to analyze.

G-quadruplex formation was analyzed using the phasor approach by Jonathan Chaires and co-workers (Buscaglia et al. 2012), titrating with KCl a solution of 2AP-labeled 22-nucleotide deoxyoligonucleotide, derived from the human telomeric repeat sequence known to form G-quadruplexes. Phasor plot analysis overcomes the problem of the complex fluorescence lifetime distributions of 2AP labelled nucleotide, because in this representation a given state of the system appears as a single point within the universal semicircle. Addition of KCl produced a multiphasic trajectory, capturing subtle changes during the folding of G-quadruplex upon $K^+$ addition.

Additionally, the authors performed a similar experiment using a 22-nucleotide human telomeric G-quadruplex forming sequence, dual-labeled with the FRET pair 6-carboxyfluorescein (donor) and tetramethylrhodamine (acceptor). The donor-only labeled samples showed no significant movement in the phasor plot and remained near the universal semicircle as the KCl concentration increased. In contrast, the dual-labeled samples exhibited a non-linear trajectory, similar to those observed in the 2AP-labeled samples. These changes, which reflect the folding of specific loop regions, were not detectable by traditional techniques such as circular dichroism.

**Concluding remarks**

Phasor methods have become increasingly popular for analyzing fluorescence microscopy data, offering a simple, model-free approach that has democratized the access to advanced imaging technologies (Malacrida 2023). Several excellent reviews have highlighted the application of phasors in fluorescence microscopy (Ranjit et al. 2018; Datta et al. 2020; Malacrida et al. 2021; Torrado et al. 2022; Garcia et al. 2023; Torrado et al. 2024; Li 2025), offering valuable insights for readers interested in these strategies. However, it is worth noting that the power of phasors extends well beyond microscopy.

In this work we have outlined the basic principles of phasor analysis and described how this approach has provided simple tools for analyzing fluorescence data across various biophysical areas. This analysis does not require postulating any underlying mechanism, but provides valuable information about the process being investigated that complements those obtained by traditional methods. Moreover, phasor analysis not only streamlines data processing but also allows the detection of some features of the phenomena under study that traditional biophysical approaches may overlook.

**Funding**

This study was funded by Universidad de Buenos Aires (UBACyT 306BA), CONICET (PIP 3266CO), Chan Zuckerberg Initiative DAF (2020-225439, 2021-240122 and 2022-252604) and Fondo para la Convergencia Estructural del Mercosur (COF 03/11).

**Conflict of interest**

The authors declare no competing interests.

**References**

Alcala JR, Gratton E, Prendergast FG (1987) Interpretation of fluorescence decays in proteins using continuous lifetime distributions. Biophys J 51:925-936. https://doi.org/10.1016/S0006-3495(87)83420-3

Araújo AEA, Tonidandel DAV (2013) Steinmetz and the Concept of Phasor: A Forgotten Story. J Control Autom Electr Syst 24:388–395. https://doi.org/10.1007/s40313-013-0030-5

Bader AN, Visser NV, van Amerongen H, Visser AJWG (2014) Phasor approaches simplify the analysis of tryptophan fluorescence data in protein denaturation studies. Methods Appl Fluoresc. 2:045001. http://doi.org/10.1088/2050-6120/2/4/045001

Berberan-Santos MN (2001). Pioneering Contributions of Jean and Francis Perrin to Molecular Luminescence. In: Valeur, B., Brochon, JC. (eds) New Trends in Fluorescence Spectroscopy. Springer Series on Fluorescence, vol 1. Springer, Berlin, Heidelberg

Berberan-Santos MN (2015) Phasor plots of luminescence decay functions. Chem Phys 449:23-33. https://doi.org/10.1016/j.chemphys.2015.01.007

Buscaglia R, Jameson DM, Chaires JB (2012) G-quadruplex structure and stability illuminated by 2-aminopurine phasor plots. Nucl Ac Res 40:4203–4215. https://doi.org/10.1093/nar/gkr1286

Cattoni DI, Kaufman SB, González Flecha FL (2009) Kinetics and thermodynamics of the interaction of 1-anilino-naphthalene-8-sulfonate with proteins. Biochim Biophys Acta 1794: 1700-1708. https://doi.org/10.1016/j.bbapap.2009.08.007

Cattoni DI, González Flecha FL, Argüello JM (2008) Thermal stability of CopA, a polytopic membrane protein from the hyperthermophile *Archaeoglobus fulgidus*. Arch Biochem Biophys 47:198-206. https://doi.org/10.1016/j.abb.2007.12.013

Cioaba SM, Linde W (2023) A Bridge to Advanced Mathematics. From Natural to Complex Numbers. American Mathematical Society.

Creighton TE (1988) Toward a better understanding of protein folding pathways. Proc Natl Acad Sci USA 85:5082-5086. https://doi.org/10.1073/pnas.85.14.5082

Danielson D (2003) Vectors and Tensors in Engineering and Physics. CRC Press. Boca Raton. https://doi.org/10.1201/9780429502774

Datta R, Heaster TM, Sharick JT, Gillette AA, Skala MC (2020) Fluorescence lifetime imaging microscopy: fundamentals and advances in instrumentation, analysis, and applications. J Biomed Opt 25:1-43. doi: https://doi.org/10.1117/1.JBO.25.7.071203

Digman MA, Caiolfa VR, Zamai M, Gratton E (2008) The Phasor Approach to Fluorescence Lifetime Imaging Analysis. Biophys J 94:L14-L16. https://doi.org/10.1529/biophysj.107.120154

Dodes Traian MM, Cattoni DI, Levi V, González Flecha FL (2012) A two-stage model for lipid modulation of the activity of integral membrane proteins. PLoS one 7: e39255. https://doi.org/10.1371/journal.pone.0039255

Durrant AV (1996) Vectors in Physics and Engineering. CRC Press. Boca Raton. https://doi.org/10.1201/9780203734391

Eftink MR (1994) The use of fluorescence methods to monitor unfolding transitions in proteins. Biophys J 66:482-501. https://doi.org/10.1016/s0006-3495(94)80799-4

Fanali G, di Masi A, Trezza V, Marino M, Fasano M, Ascenzi P (2012) Human serum albumin: From bench to bedside. Mol Asp Med 33:209-290. https://doi.org/10.1016/j.mam.2011.12.002

Farina S, Acconcia G, Labanca I, Ghioni M, Rech I (2021) Toward ultra-fast time-correlated single-photon counting: A compact module to surpass the pile-up limit. Rev Sci Instrum. 92:063702. https://doi.org/10.1063/5.0044774

Fereidouni F, Bader AN, Gerritsen HC (2012) Spectral phasor analysis allows rapid and reliable unmixing of fluorescence microscopy spectral images Opt Express 20:12729-12741 https://doi.org/10.1364/oe.20.012729

Ferrer-Sueta G, Manta B, Botti H, Radi R, Trujillo M, Denicola A (2011) Factors affecting protein thiol reactivity and specificity in peroxide reduction. Chem Res Toxicol 24:434–450. https://doi.org/10.1021/tx100413v

Finkelstein AV, Bogatyreva NS, Ivankov DN, Garbuzynskiy SO (2022) Protein folding problem: enigma, paradox, solution. Biophys Rev 14:1255–1272. https://doi.org/10.1007/s12551-022-01000-1

García MJ, Kamaid A, Malacrida L (2023) Label-free fluorescence microscopy: revisiting the opportunities with autofluorescent molecules and harmonic generations as biosensors and biomarkers for quantitative biology. Biophys Rev 15:709–719. https://doi.org/10.1007/s12551-023-01083-4

Gaviola E (1927) Ein Fluorometer. Apparat zur Messung von Fluoreszenzabklingungszeiten. Z Phys 42:853-861. https://doi.org/10.1007/BF01776683

Gratton E (2016) Measurements of fluorescence decay time by the frequency domain method. In Perspectives on Fluorescence: A Tribute to Gregorio Weber. Jameson DM (Ed) Springer, Cham. https://doi.org/10.1007/4243_2016_15

Gunther G, Malacrida L, Jameson D, Gratton E, Sanchez SA (2021) LAURDAN since Weber: The Quest for Visualizing Membrane Heterogeneity. Acc Chem Res 54, 4 https//doi.org/10.1021/acs.accounts.0c00687

Gutfreund H. (1995) Kinetics for the Life Sciences: Receptors, Transmitters and Catalysts. Cambridge University Press. London. https://doi.org/10.1017/CBO9780511626203

Hanley QS, Clayton AH (2005) AB-plot assisted determination of fluorophore mixtures in a fluorescence lifetime microscope using spectra or quenchers. J Microsc 218:62–67. https://doi.org/10.1111/j.1365-2818.2005.01463.x

James NG, Ross J A, Stefl M, Jameson DM (2011) Applications of phasor plots to in vitro protein studies. Anal Biochem 410:70–6. https://doi.org/10.1016/j.ab.2010.11.011

Jameson DM (1998) Gregorio Weber, 1916–1997: A fluorescent lifetime. Biophys J 75:419–421. https://doi.org/10.1016/S0006-3495(98)77528-9

Jameson DM (2001) The seminal contributions of Gregorio Weber to Modern Fluorescence Spectroscopy. In New Trends in Fluorescence Spectroscopy, Eds.B. Valeur and J.-C. Brochon, Springer-Verlag, Berlin, pp. 35–58. https://doi.org/10.1007/978-3-642-56853-4_3

Jameson DM (2014) Introduction to Fluorescence. Taylor & Francis, Boca Raton. https://doi.org/10.1201/b16502

Jameson DM, Gratton E and Hall RD (1984) The measurement phase and modulation fluorometry. Appl Spectrosc Rev 20:55–106. https://doi.org/10.1080/05704928408081716

Kolesnikova S, Curtis EA (2019) Structure and Function of Multimeric G-Quadruplexes. Molecules 24:3074. https://doi.org/10.3390/molecules24173074

Li D, Liu X, Dong F, Li W (2025) Advancements in phasor-based FLIM: multi-component analysis and lifetime probes in biological imaging. J Mater Chem B 13:472-484. https://doi.org/10.1039/d4tb01669f

Lippitz M, Kulzer F, Orrit M (2005) Statistical evaluation of single nano-object fluorescence. Chemphyschem. 6:770-89. https://doi.org/10.1002/cphc.200400560

Liu H, Sun S, Zheng L, Wang G, Tian W, Zhang D, Han H, Zhu J, Wei Z (2021) Review of laser-diode pumped Ti:sapphire laser. Microw Opt Technol Lett 63: 2135–2144. https://doi.org/10.1002/mop.32882

Malacrida L, Gratton E, Jameson DM, (2015) Model-free methods to study membrane environmental probes: a comparison of the spectral phasor and generalized polarization approaches. Methods Appl Fluoresc 3:047001. https://doi.org/10.1088/2050-6120/3/4/047001

Malacrida L, Ranjit S, Jameson DM, Gratton E (2021) The Phasor Plot: A universal circle to advance fluorescence lifetime analysis and interpretation. Ann Rev Biophys 50:575-593 https://doi.org/10.1146/annurev-biophys-062920-063631

Malacrida L (2023) Phasor plots and the future of spectral and lifetime imaging. Nature Methods 20:965-967.https://doi.org/10.1038/s41592-023-01906-y

Mangiarotti A, Bagatolli LA (2021) Impact of macromolecular crowding on the mesomorphic behavior of lipid self-assemblies. Biochim Biophys Acta 1863:183728. https://doi.org/10.1016/j.bbamem.2021.183728

Minsky A (2004) Information Content and Complexity in the High-Order Organization of DNA. Ann Rev Biophys 33:317-342 https://doi.org/10.1146/annurev.biophys.33.110502.133328

Monsen RC, Trent JO, Chaires JB (2022) G-quadruplex DNA: A Longer Story. Acc Chem Res 55:3242–3252. https://doi.org/10.1021/acs.accounts.2c00519

Montecinos-Franjola F, James NG, Concha-Marambio L, Brunet JE, Lagos R, Monasterio O, Jameson DM (2014) Single tryptophan mutants of FtsZ: Nucleotide binding/exchange and conformational transitions. Biochim Biophys Acta 1844:1193-1200. http://dx.doi.org/10.1016/j.bbapap.2014.03.012

Moon CP, Fleming KG (2011) Using tryptophan fluorescence to measure the stability of membrane proteins folded in liposomes Meth Enzymol 492:189-211 https://doi.org/10.1016/B978-0-12-381268-1.00018-5

Neely RK, Daujotyte D, Grazulis S, Magennis SW, Dryden DTF, Klimašauskas S, Jones AC (2005) Time-resolved fluorescence of 2-aminopurine as a probe of base flipping in M.HhaI–DNA complexes, Nucl Ac Res 33: 6953–6960. https://doi.org/10.1093/nar/gki995

Onuchic JN, Wolynes PG (2004) Theory of protein folding. Curr Op Struct Biol 14:70-75. https://doi.org/10.1016/j.sbi.2004.01.009

Parasassi T, Gratton E (1995) Membrane lipid domains and dynamics as detected by Laurdan fluorescence. J Fluoresc 5:59–69 https://doi.org/10.1007/BF00718783

Parasassi T, Gratton E, Yu W M, Wilson P and Levi M (1997) Two-photon fluorescence microscopy of Laurdan generalized polarization domains in model and natural membranes. Biophys J 72:2413–29. https://doi.org/10.1016/S0006-3495(97)78887-8

Perez Socas LB, Ambroggio EE (2020) The influence of myristoylation, liposome surface charge and nucleic acid interaction in the partition properties of HIV-1 Gag-N-terminal peptides to membranes. Biochim Biophys Acta 1862:183421. https://doi.org/10.1016/j.bbamem.2020.183421

Perez Socas LB, Ambroggio EE (2022) Introducing the multi-dimensional spectral phasors: a tool for the analysis of fluorescence excitation-emission matrices. Methods Appl Fluoresc 10: 025003. https://doi.org/10.1088/2050-6120/ac5389

Perez Socas LB, Valdivia-Pérez JA, Fanani ML, Ambroggio EE (2024) Multidimensional Spectral Phasors of LAURDAN's Excitation–Emission Matrices: The Ultimate Sensor for Lipid Phases? J Am Chem Soc 146:17230–17239. https://doi.org/10.1021/jacs.4c03443

Perrais D (2022) Cellular and structural insight into dynamin function during endocytic vesicle formation: a tale of 50 years of investigation. Biosci Rep 42:BSR20211227. https://doi.org/10.1042/BSR20211227

Ptitsyn OB (1995) Molten globule and protein folding. Adv Prot Chem 47:83-229. https://doi.org/10.1016/S0065-3233(08)60546-X

Rajasekaran N, Kaiser CM (2024) Navigating the complexities of multi-domain protein folding. Curr Op Struct Biol 86:102790. https://doi.org/10.1016/j.sbi.2024.102790

Ranjit S, Malacrida L, Jameson DM, Gratton E (2018) Fit-free analysis of fluorescence lifetime imaging data using the phasor approach. Nat Protoc 13:1979–2004. https://doi.org/10.1038/s41596-018-0026-5

Recoulat Angelini AA, Placenti MA, Melian NA, Sabeckis ML, Burgardt NI, González-Lebrero RM, Roman EA, González Flecha FL (2021) Cu(I)-transport ATPases. Molecular architecture, catalysis and adaptation to extreme environments. Adv Med Biol 180:65-130

Recoulat Angelini AA, Roman EA, González Flecha FL (2024) The structural stability of membrane proteins revisited: Combined thermodynamic and spectral phasor analysis of SDS-induced denaturation of a thermophilic Cu(I)-transport ATPase. J Mol Biol 436:168689. https://doi.org/10.1016/j.jmb.2024.168689

Redford GI, Clegg RM (2005) Polar plot representation for frequency-domain analysis of fluorescence lifetimes. J Fluoresc 15:805-815. https://doi.org/10.1007/s10895-005-2990-8

Reinhart GD, Marzola P, Jameson DM, Gratton E (1991). A method for on-line background subtraction in frequency domain fluorometry. J Fluoresc 1:153–162. https://doi.org/10.1007/BF00865362

Roman EA, González Flecha FL (2014) Kinetics and Thermodynamics of Membrane Protein Folding. Biomolecules 4:354-373. https://doi.org/10.3390/biom4010354

Roman EA, Argüello JM, González Flecha FL (2010) Reversible unfolding of a thermophilic membrane protein in phospholipid/detergent mixed micelles. J Mol Biol 397:550-559. https://doi.org/10.1016/j.jmb.2010.01.045

Ross JA, Jameson DM (2008) Time-resolved methods in biophysics. 8. Frequency domain fluorometry: applications to intrinsic protein fluorescence. Photochem Photobiol Sci 11:1301-12. https://doi.org/10.1039/b804450n

Royer C (2006) Probing Protein Folding and Conformational Transitions with Fluorescence. Chem Rev 106:1769–1784. https://doi.org/10.1021/cr0404390

Shelby SA, Veatch SL (2023) The Membrane Phase Transition Gives Rise to Responsive Plasma Membrane Structure and Function. Cold Spring Harb Perspect Biol 15:a041395. https://doi.org/10.1101/cshperspect.a041395

Siddiqui S, Ameen F, Rehman Su, Sarwar T, Tabish M (2021) Studying the interaction of drug/ligand with serum albumin. J Mol Liq 336:116200. https://doi.org/10.1016/j.molliq.2021.116200

Silber N, Matos de Opitz CL, Mayer C, Sass P (2020) Cell Division Protein FtsZ: From Structure and Mechanism to Antibiotic Target. Future Microbiol 15:801-831. https://doi.org/10.2217/fmb-2019-0348

Spencer RD, Weber G (1969). Measurements of subnanosecond fluorescence lifetimes with a cross-correlation phase fluorimeter. Ann N Y Acad Sci 158:361-376. https://doi.org/10.1111/j.1749-6632.1969.tb56231.x

Stefl M, James NG, Ross JA, Jameson DM (2011) Applications of phasors to in vitro time-resolved fluorescence measurements. Anal Biochem 410:62–9. https://doi.org/10.1016/j.ab.2010.11.010

Steinmetz CP (1893) Complex Quantities and Their Use in Electrical Engineering. Proceedings of the International Electrical Congress.

Stokes GG (1852) On the change in refrangibility of light. Phil Trans Royal Soc London 142:463–562. https://doi.org/10.1098/rstl.1852.0022

Teale FWJ, Weber G (1957) Ultraviolet fluorescence of the aromatic amino acids. Biochem J 53:476-482. https://doi.org/10.1042/bj0650476

Torrado B, Malacrida L, Ranjit S (2022) Linear Combination Properties of the Phasor Space in Fluorescence Imaging. Sensors 22:999. https://doi.org/10.3390/s22030999

Torrado B, Pannunzio B, Malacrida L, Digman MA (2024) Fluorescence lifetime imaging microscopy. Nat Rev Methods Primers 4:80. https://doi.org/10.1038/s43586-024-00358-8

Valeur B, Berberan-Santos MN (2011) A Brief History of Fluorescence and Phosphorescence before the Emergence of Quantum Theory. J Chem Ed 88:731–738. https://doi.org/10.1021/ed100182h

Valeur B, Berberan-Santos MN (2012) Molecular Fluorescence: Principles and Applications. Wiley‐VCH Verlag GmbH & Co. KGaA. https://doi.org/10.1002/9783527650002

Vallée-Bélisle A, Michnick S (2012) Visualizing transient protein-folding intermediates by tryptophan-scanning mutagenesis. Nat Struct Mol Biol 19:731–736. https://doi.org/10.1038/nsmb.2322

Veatch SL, Keller SL (2003) Separation of Liquid Phases in Giant Vesicles of Ternary Mixtures of Phospholipids and Cholesterol. Biophys J 85:3074–3083. https://doi.org/10.1016/S0006-3495(03)74726-2

Villar SF, Dalla-Rizza J, Möller MN, Ferrer-Sueta G, Malacrida L, Jameson DM, Denicola A. (2022) Fluorescence Lifetime Phasor Analysis of the Decamer–Dimer Equilibrium of Human Peroxiredoxin 1. Int J Mol Sci 23:5260. https://doi.org/10.3390/ijms23095260

Villar SF, Möller MN, Denicola A (2023) Biophysical tools to study the oligomerization dynamics of Prx1-class peroxiredoxins. Biophys Rev 15:601–609. https://doi.org/10.1007/s12551-023-01076-3

Weber G (1952) Polarization of the fluorescence of macromolecules. II. Fluorescent conjugates of ovalbumin and bovine serum albumin, Biochem J 51:155–167. https://doi.org/10.1042/bj0510155

Weber G (1960) Enumeration of components in complex systems by fluorescence spectrophotometry. Nature 190:27–29. https://doi.org/10.1038/190027a0

Weber G (1981) Resolution of the fluorescence lifetimes in a heterogeneous system by phase and modulation measurements. J Phys Chem 85:949–53. https://doi.org/10.1021/j150608a006

Weber G (1992) Protein Interactions, Chapman, New York.

Weber G, Farris FJ (1979) Synthesis and Spectral Properties of a Hydrophobic Fluorescent Probe: 6-Propionyl-2-(dimethylamino)- naphthalene. Biochemistry 18:3075−3078. https://doi.org/10.1021/bi00581a025

Weber G, Laurence DJR (1954) Fluorescent indicators of adsorption in aqueous solution and on the solid phase. Biochem J 56: xxxi

Weber G, Teale FWJ (1959) Polarization of the ultraviolet fluorescence and electronic energy transfer in proteins. Biochem J 72:15p